\documentclass[prb,preprint]{revtex4}
\usepackage {graphicx}

\begin{document}

\title{Simple derivation of Young, Wenzel and Cassie-Baxter equations and its interpretations.}

\author{S. Banerjee\footnote{Email: sangam.banerjee@saha.ac.in}}

\address{Surface Physics Division,  Saha Insitute of Nuclear Physics, 1/AF Bidhannagar, Kolkata 700 064, India}

\begin{abstract} 
In this paper we have derived Young's, Wenzel's and  Cassie-Baxter's equations using conceptual model rather than showing rigorous derivation to help the new-comers in this field. We then pointed out that if the substrate is initially hydorphilic then one can modify the surface morphology and make the substrate to become hydrophobic or superhydrophobic. But, if the substrate is initially hydrophobic then one can only make it superhydrophobic but not hydrophilic by modifying the surface morphology using the formalisms mentioned in this paper.
\end{abstract}

\maketitle

Understanding wetting of liquid on a solid surface at micro and nano level will be very important from the point of view of various biological and physical applications. Wetting of a liquid on a solid surface depends not only on the chemical nature but also on the morphological structure of their surfaces \cite{Quere}. Here, we would like to show simple derivations of Young's \cite{Young}, Wenzel's \cite{Wenzel} and  Cassie-Baxter's \cite{CB} equations to quantitatively obtain the contact angles as a function of various surface morphology. We would derive the results conceptually rather than showing rigorous derivation to help the new-comers in this field. The change in the surface morphology modifies the surface and the interfacial energy of the solid surface and hence its wetting property. To begin with, when a drop of liquid "L" is put on top of a solid substrate "S" in an environment of vapor "V", the equilibrium shape of the liquid is obtained when the total free energy of the surfaces of the liquid in contact with the solid and vapour is minimum. The free surface energy for solids and liquids are expressed in terms of $\gamma$'s. These $\gamma$'s are the change in the free energy of the medium when the surface area of the medium is increased by unit area. $\gamma$ for solids are expressed in energy per unit area (surface energy) and for liquids it is expressed in tension per unit length (surface tension), both these units are dimensionally same. A liquid drop comes in contact and adheres with the solid surface forming a finite contact area if and only if the total energy of the system reduces i.e., reduction of the total energy of the system by the amount of the work of adhesion of the liquid to the solid. Thus, the work of adhesion per unit area $w_{a,SL}$ is the amount of energy per unit area required to seperate the liquid and the solid in a vapor medium. Hence, when a liquid comes in contact with the solid, the work of adhesion per unit area is expressed as \cite{Israel}:

\begin{equation}
w_{a,SL} = \gamma_{SV}+ \gamma_{LV} - \gamma_{SL} \hspace{2cm}per \hspace{0.2cm} unit\hspace{0.2cm} area.
\end{equation}
and hence,
\begin{equation}
\gamma_{SL}= \gamma_{SV}+ \gamma_{LV} - w_{a,SL} \hspace{2cm}per \hspace{0.2cm} unit\hspace{0.2cm} area.
\end{equation}

\noindent where $\gamma_{LV}$, $\gamma_{SV}$ and $\gamma_{SL}$ are the surface energy (surface tension) of the liquid/vapor (LV) interface, of the solid/vapor (SV) interface and of the solid/liquid (SL) interface respectively. When the work of adhesion $w_{a,SL}$ ($w_{a,SL} =  \gamma_{SV} + \gamma_{LV} - \gamma_{SL}$) is positive then there is an attraction between the solid and liquid leading to wetting and when it is negative then there is a repulsion between them leading to non-wetting of the liquid on the solid substrate. When the liquid drop is in vapor, its total interfacial free energy is (see fig. 1(a)) \cite{Israel}:

\begin{equation}
W_{tot,liquid} = \gamma_{LV}A_{sphere}
\end{equation}

\noindent $A_{sphere}$ is the surface area of the liquid sphere. When the liquid is in contact with the solid substrate then the total area of the liquid drop is equal to ($A_{SL} +A_{LV}$), where $A_{SL}$ and $A_{LV}$ are the area of the liquid drop at the solid/liquid (SL) and the liquid/vapor (LV) interface respectively (see fig. 1(b and c)). From fig. 1(c) we obtain:

\begin{equation}
W_{tot,liquid} = \gamma_{LV}(A_{SL}+A_{LV}) - w_{a,SL} A_{SL}
\end{equation}

\noindent At equilibrium, the surface energy of the liquid is minimised by minimizing the contact areas i.e., 

\begin{equation}
\gamma_{LV}(dA_{SL}+dA_{LV}) - w_{a,SL} dA_{SL} = 0
\end{equation}

\noindent and assuming $dA_{LV}/dA_{SL} = cos\theta_Y$ (see fig. 1(d)) one obtains the famous Young's equation \cite{Young} for contact angle in terms of the surface tensions i.e., 

\begin{equation}
cos\theta_Y = (\gamma_{SV} - \gamma_{SL})/\gamma_{LV}
\end{equation}
or 
\begin{equation}
\gamma_{SL} = \gamma_{SV} - \gamma_{LV}cos\theta_Y
\end{equation}

For $\gamma_{SV} > \gamma_{SL}$ then $0^o < \theta_Y < 90^o$ and for $\gamma_{SL} > \gamma_{SV}$ then $90^o < \theta_Y < 180^o$. From the Young's equation one can see that the interfacial tension between the solid and liquid $\gamma_{SL}$ is lower than $\gamma_{SV}$ only when $\theta_Y < 90^o$, this happens only in the case of wetting. If $\gamma_{SL} > \gamma_{SV}$ which can happen only when $\theta_Y > 90^o$ considering $\gamma_{LV}$ is always finite and positive value, then to minimize the total surface/interfacial energy of the liquid, the area of contact of the liquid with the solid surface ($A_{SL}$) will be reduced and in this case the liquid behaves as a non-wetting. Thus the contact angle depends on the optimization of the area of contact of the solid/liquid and the liquid/vapor interface. Thus, the wetting property of a liquid on a flat solid substrate can be understood from the thermodynamics of surfaces by evaluating the work of adhesion leading to Young's equation. Another simple derivation of Young's equation can be obtained by balancing the forces at the line of contact where all the three medium (solid, liquid and vapor) meets as shown schematically in fig. 2(a,b). Thus, the equilibrium contact angle the liquid drop makes with the smooth and flat substrate depends on the value of the difference between $\gamma_{SV}$ and $\gamma_{SL}$ as expressed above.

It is interesting to note that the micro and nano structures on the surface can modify the wetting property of the solid surface. This effect was pointed out by Wenzel \cite{Wenzel}. The Wenzel case is shown schematically in fig. 2(c). Wenzel considered rough surface and characterized it by a roughness ratio factor $"r"$ defined as the ratio of the true area of the solid surface to its projection i.e., $r = A_{rough}/A_{flat}$ and $r$ is always greater than 1. Thus as $r$ increases the total surface/interface energy also increases because the true area increases. Substituting the roughness ratio factor $"r"$ in Young's equation ($\gamma_{SV}$ becomes $r\gamma_{SV}$ and $\gamma_{SL}$ becomes $r\gamma_{SL}$), one can obtain the Wenzel's equation i.e., 

\begin{equation}
cos\theta_W = r (\gamma_{SV} - \gamma_{SL})/\gamma_{LV} = r cos\theta_Y 
\end{equation}

\noindent where $\theta_Y$ is the contact angle for the flat substrate. Wenzel's equation immediately suggests that if $\theta_Y < \pi/2$ then $\theta_W < \theta_Y$ and hence in this case introduction of roughness will enhance the tendency for liquid to wet further and on the other hand if $\theta_Y > \pi/2$ then $\theta_W > \theta_Y$ and in this case the de-wetting tendency will be enhanced as shown in fig. 3 indicated by arrows. Thus in Wenzel's model, the roughness can enhance the wetting property of the solid surface further if the liquid was initially wetting on flat(smooth) surface (enhancing the hydrophilicity) and on the other hand if inititally the liquid was non-wetting on the flat surface, then the introduction of the roughness can further make the surface of the solid substrate non-wetting, leading to enhancement of hydrophobicity. 

Later, Cassie and Baxter \cite{CB} (CB) further considered that if the substrate is flat but consists of randomly distributed $"n"$ different type of materials on the surface and each of these materials are characterized by their own surface energies/tensions i.e., $\gamma_{i,SL}$ and $\gamma_{i,SV}$ with the respective material fraction $f_i$ on the substrate surface with $f_1 + f_2 + ... + f_n = 1$, $\gamma_{SV} = \Sigma_i^nf_i(\gamma_{i,SV})$ and $\gamma_{SL} = \Sigma_i^nf_i(\gamma_{i,SL})$ and substituting in Young's equation we obtain a modified contact angle expressed as: 

\begin{equation}
cos\theta_{CB} = \Sigma_i^nf_i(\gamma_{i,SV} - \gamma_{i,SL})/\gamma_{LV} = \Sigma_i^nf_i cos\theta_{i,Y} 
\end{equation}

\noindent which is known as the Cassie-Baxter (CB) equation. In certain type of surfaces with a specific texture or morphology, air can be trapped in between the asperities like in a lotus leaf "lotus effect", such that the liquid drop sits on a surface having distribution of solid asperities and air-pockets (two component surface material) and their fractions are $f_S$ and $f_V$ respectively where $f_S + f_V = 1$. Substituting in the CB equation for the solid-liquid fraction $f_1 = f_S$ and $cos\theta_{1,Y} = cos\theta_Y$ and for the air-pocket fraction $f_2 = f_V$ and $cos\theta_{2,Y} = -1$ because the fraction is completely dry (non-wetting) and combining the roughness ratio factor $"r"$ with the Cassie-Baxter equation, we get 

\begin{equation}
cos\theta_{CB} = rf_S cos\theta_Y - f_V = rf_S cos\theta_Y + f_S -1 
\end{equation}

By looking at the Cassie-Baxter equation we are tempted to tailor the surface topography such that the solid fraction $f_S$ is made to approach zero by introducing aspereties on the surface, so that $\theta_{CB}$ approaches 180$^o$ i.e., complete non-wetting as shown schematically in fig. 2(d). 

In fig. 3 we have plotted $\theta_{CB}$ versus $\theta_Y$ and we observe that $\theta_{CB}$ is always greater than or equal to $\theta_W$ and as the value of $f_S$ decreases the value of $\theta_{CB}$ increases. In fig. 4 we have plotted the Cassie-Baxter angle $\theta_{CB}$ as a function of the roughness ratio factor $"r"$ and we observe that when $\theta_Y$ is less than 90$^o$ ie., when the smooth and flat substrate is hydrophilic, $\theta_{CB}$ can cross over from hydrophobic to hydrophilic upon increasing the value of $"r"$. We also observe that for the same value of $"r"$ as $f_S$ decreases the contact angle $\theta_{CB}$ increases. An interesting point to note that if $\theta_Y$ is greater than 90$^o$ ie., when the smooth and flat substrate is hydrophobic, $\theta_{CB}$ "$can not$" cross over from hydrophobic to hydrophilic on increasing the value of $"r"$ unlike in the previous situation. Thus this indicates that by merely modifying the surface morphology of any smooth and flat hydrophobic substrate can not be modified to become hydrophilic using the above Wenzel and Cassie-Baxter formalism. But recently \cite{Sangam} it has been shown experimentally that if substrate have undulation then for certain undulation length scale the apparent contact angle is less than 90$^o$ i.e., appears to be wetting even with the Cassie-Baxter like asperities riding over the undulation. Now one question arises: can one distinguish Wenzel state from a Cassie-Baxter state? Yes, a drop in the Wenzel state will have a high sliding angle than the Cassie-Baxter state which have low sliding angle and hence in the CB state the drop will easily roll off on slight tilting of the substrate.

To, summarise, we have shown simple derivations of Young's, Wenzel's and Cassie-Baxter's equations for wetting of smooth and rough surfaces of a solid. However, we have pointed out that if the substrate is initially hydorphilic then one can modify the surface morphology and make the substrate to become hydrophobic or super hydrophobic. But, if the substrate is initially hydrophobic then one can only make it superhydrophobic but not hydrophilic using the formalisms mentioned in this paper.

\newpage

\begin{figure}
\includegraphics*[width=12.0cm,angle=270]{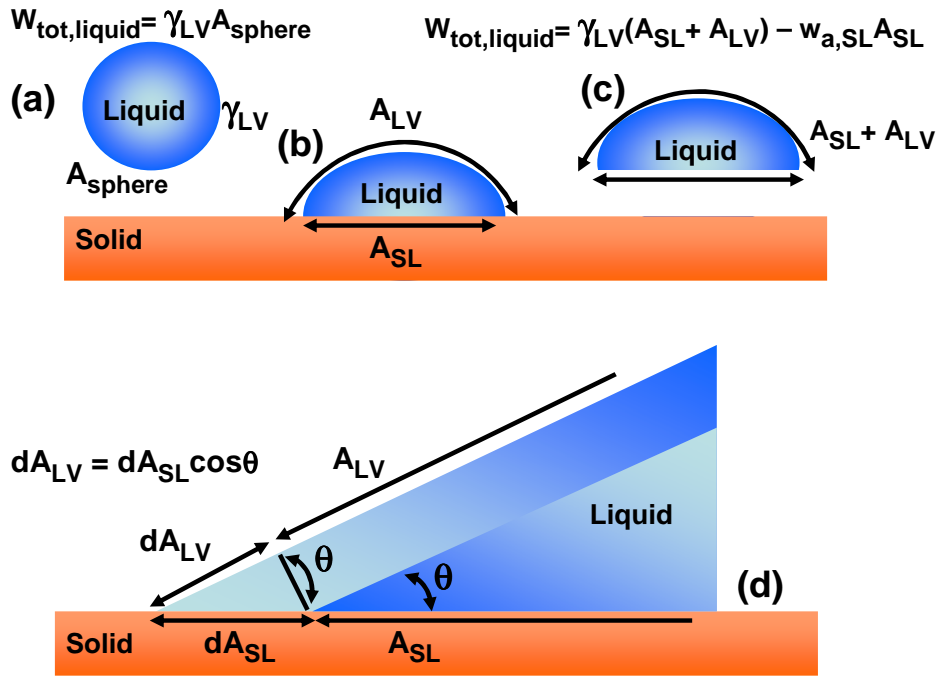}
\caption{(Colour online) (a) Area of liquid drop is $A_{sphere}$, (b) Curved area of the liquid drop at liquid/vapor interface is $A_{LV}$ and the flat area of the liquid drop in contact with the solid substrate is $A_{SL}$, (c) Total interfacial area of the deformed liquid drop is $A_{LV}+A_{SL}$ and (d) Geometrically showing $dA_{LV}/dA_{SL} = cos\theta$}
\end{figure}

\begin{figure}
\includegraphics*[width=12.0cm,angle=270]{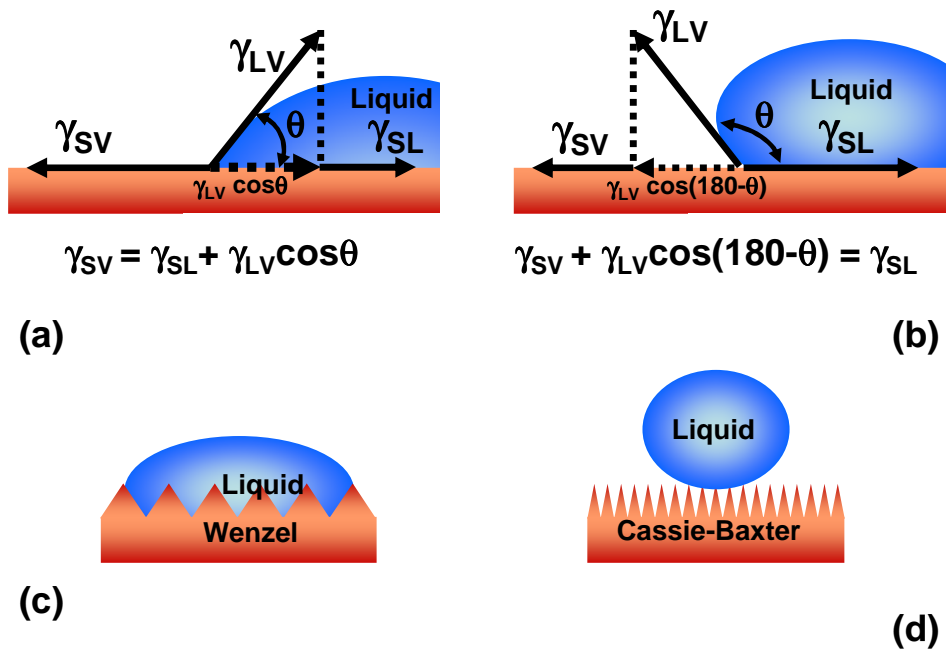}
\caption{(Colour online)(a,b) Schematically we show a simple derivation of Young's equation using surface tension vectors for a liquid on a solid substrate.(c) Wenzel's model (d) Cassie-Baxter's model }
\end{figure}

\begin{figure}
\includegraphics*[width=12.0cm,angle=270]{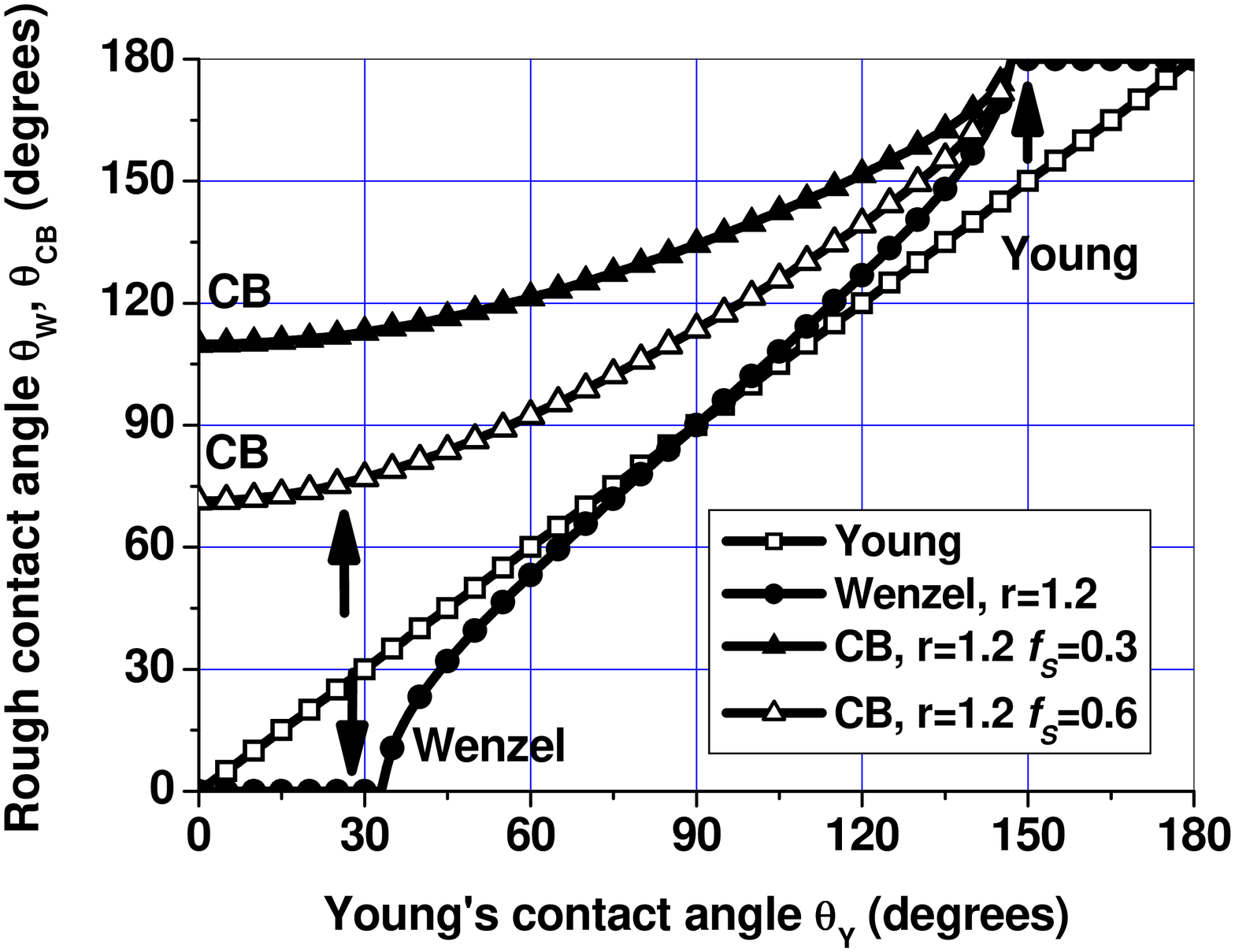}
\caption{Evolution of contact angle from Wenzel's model (with $r$ = 1.2) and Cassie-Baxter's model (with $r$ = 1.2 and $f_S$ = 0.3 and 0.6) with respect to Young's contact angle (flat substrate)}
\end{figure}

\begin{figure}
\includegraphics*[width=12.0cm]{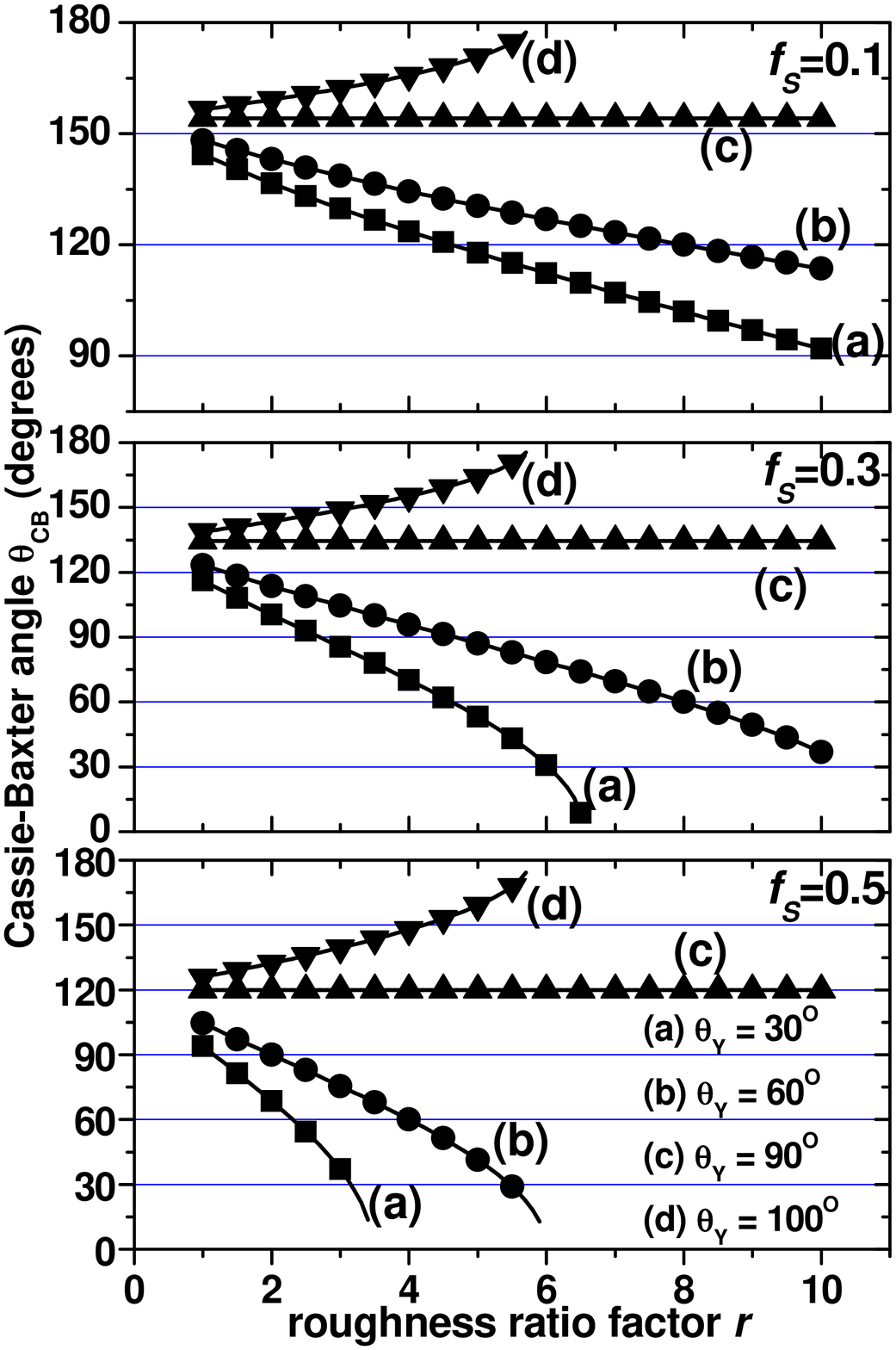}
\caption{Evolution of contact angle from Cassie-Baxter's model as a function of $r$, $f_S$ and Young's contact angle $\theta_Y$. We observe that for $\theta_Y$ $\geq$ 90$^o$ modification of surface morphology by changing roughness ratio factor $r$ and $f_S$ does not modify the Cassie-Baxter angle $\theta_{CB}$ to less than $90^o$ i.e., the surface cannot be made hydrophilic if initially for the flat substrate the surface is hydrophobic.}
\end{figure}

\end{document}